\documentclass{article}
\usepackage{times}
\usepackage{amsfonts}
\usepackage{graphicx}
\begin{document}
\noindent
{\Large  ENTROPY, TOPOLOGICAL THEORIES AND\\ EMERGENT QUANTUM MECHANICS}\\

\vskip1cm
\noindent
{\bf D. Cabrera}${}^{a}$, {\bf P. Fern\'andez de C\'ordoba}${}^{b}$, 
{\bf  J.M. Isidro}${}^{c}$ and {\bf J. Vazquez Molina}${}^{d}$\\
Instituto Universitario de Matem\'atica Pura y Aplicada,\\ Universidad Polit\'ecnica de Valencia, Valencia 46022, Spain\\
${}^{a}${\tt dacabur@upvnet.upv.es}, ${}^{b}${\tt pfernandez@mat.upv.es}\\
${}^{c}${\tt joissan@mat.upv.es}, ${}^{d}${\tt joavzmo@doctor.upv.es}\\
\vskip.5cm
\noindent
{\bf Abstract} The classical thermostatics of equilibrium processes is shown to possess a quantum mechanical dual theory with a finite dimensional Hilbert space of quantum states. Specifically, the kernel of a certain Hamiltonian operator becomes the Hilbert space of quasistatic quantum mechanics. The relation of thermostatics to topological field theory is also discussed in the context of the approach of emergence of quantum theory, where the concept of entropy plays a key role.

\tableofcontents

\section{Motivation}\label{gruende}

The approach of emergence of quantum mechanics  has provided interesting clues into the deeper structure of the theory. The statement that {\it standard quantum mechanics is an emergent phenomenon}\/ \cite{ELZE1, ELZE2, THOOFT} has found further support in a series of papers, some of which have been reviewed in ref. \cite{VALENCIA}. Although this is a huge topic to summarise here, let us briefly mention some key points of this approach. The  underlying notion is that it provides a coarse grained version of some deeper theory, out of which quantum mechanics emerges as a kind of effective description. This effective description, in using variables that arise as averages over large collections of individual entities carrying the truly fundamental degrees of freedom, ignores the underlying fine structure. 

This state of affairs is reminiscent of the relation between thermodynamics (as an emergent phenomenon) and statistical mechanics (the corresponding underlying theory). Based on this analogy, we have in previous publications (see ref. \cite{VALENCIA} and refs. therein) established a bijective map that one can define between quantum mechanics, on the one hand, and the classical thermodynamics of irreversible processes, on the other \cite{ONSAGER, PRIGOGINE}.  It must be stressed that the classical thermodynamics of irreversible processes \cite{ONSAGER, PRIGOGINE} is conceptually quite different from the usual {\it thermostatics of equilibrium}\/ as presented in the standard textbooks \cite{CALLEN}. Specifically, in the theory of irreversible processes, the continual production of entropy provides a rationale for the dissipation, or information loss, that has been argued to lie at the heart of quantum mechanics \cite{THOOFT}. The relevance of thermodynamical concepts to quantum theory and gravity has been emphasised recently in refs. \cite{BAEZ, MATONE, PADDY, PENROSE,  SVOZIL}.

It might thus appear that the usual quasistatic thermodynamics \cite{CALLEN}, {\it i.e.}\/, the thermostatics of equilibrium processes, possesses no quantum mechanical dual theory at all. In this letter we point out that such a conclusion is not true: the thermostatics of equilibrium processes {\it does}\/ have a quantum mechanical dual, namely, {\it a quasistatic quantum mechanics}\/. Under {\it quasistatic}\/ we mean that the kinetic term in the mechanical Lagrangian can be neglected compared to the potential term.

Neglecting the kinetic term in the Lagrangian function forces one to look elsewhere for the dissipative mechanism that is characteristic of quantum theory \cite{THOOFT}. In particular, such a mechanism can no longer be identified with the continual production of entropy associated with Onsager's kinetic term $L_{ij}\dot q^i\dot q^j$. The reciprocity theorem \cite{ONSAGER} ensures $L_{ij}=L_{ji}$, and dissipation requires that this matrix be positive definite; the latter two properties ensure that $L_{ij}$ qualifies as a metric. The result of neglecting the kinetic term in the Lagrangian is a mechanics bearing some resemblance to topological field theory \cite{TOPO}. Indeed, once the metric represented by the kinetic term is neglected, correlation functions  can no longer be metric dependent. Hence, while correlators can still depend on the topology of the underlying manifold, they can no longer depend on its metric structure. In our case the underlying manifold will be given by the equipotential submanifolds (within configuration space) of the potential function.

\section{A quasistatic mechanics}\label{greunda}

A quasistatic mechanics is obtained by neglecting the kinetic term $K$ in the mechanical Lagrangian $L=K-U$, and keeping only the potential term $U$:
\begin{equation}
L=-U.
\label{kwasi}
\end{equation}
Since our Lagrangian does not depend on the velocities $\dot q$, this phase space is constrained by the requirement that all momenta vanish, $p=0$, and the Hamiltonian equals
\begin{equation}
H=U.
\label{mija}
\end{equation}

We can now construct the reduced phase space corresponding to this reduced configuration space, and eventually quantise it.\footnote{For our purposes it will not be necessary to apply Dirac's theory of constrained quantisation \cite{DIRAC}.} When moving along equipotential submanifolds, the particle is effectively free; whenever motion takes place between neighbouring equipotentials, forces will cause the particle's kinetic energy to increase or decrease. However, the allowed motions must be quasistatic, so even for these motions $K$ must be negligible compared to $U$. In classical mechanics, motion along equipotential submanifolds, plus a vanishing kinetic energy, imply that a classical particle must forever stay at rest. Quantum mechanically, due to the uncertainty principle, a (more or less localised) free particle always carries a nonzero kinetic energy. So neglecting the kinetic energy of a quantum particle implies a large uncertainty in the position. This large uncertainty is reflected in a large spread of the corresponding wavepacket: the latter encompasses a large interval of different classically allowed positions, or states, all of which coalesce into a single quantum state. It is only in the limit of complete delocalisation in space that a quantum particle can carry zero kinetic energy.

We have just described an information loss mechanism whereby different classical states (different spatial positions on an equipotential submanifold, corresponding to different classically allowed equilibrium states) are lumped together into just one quantum state. This information loss has been argued to be a key feature of the quantum world.

\section{The thermostatics dual to quasistatic mechanics}\label{duegreuende}

We claim that {\it the quasistatic quantum mechanical model described in section \ref{greunda} possesses a dual theory: the classical  thermostatics of equilibrium processes}\/. In what follows we will exhibit the claimed duality explicitly.

The classical thermostatics of equilibrium \cite{CALLEN} is a theory of quasistatic processes. In particular, all kinetic energies are neglected; the processes described either are in thermal equilibrium, or at most differ infinitesimally from thermal equilibrium. This feature is in sharp contrast with the thermodynamics of irreversibility \cite{ONSAGER, PRIGOGINE}, that we described in previous publications  \cite{VALENCIA} as a thermodynamical dual of quantum mechanics, {\it whenever the kinetic energies involved could not be neglected}\/.

Next we recall that classical thermostatics is, like quantum mechanics, an emergent theory. By {\it emergent}\/ we mean that classical thermostatics is the result of coarse graining over very many microscopic degrees of freedom; the resulting theory renounces the knowledge of detailed information about its constituent degrees of freedom, retaining just a handful of relevant averages such as pressure, volume and temperature. In other words, {\it an information loss mechanism is at work}\/. This situation is similar to that described in section \ref{greunda} for the passage from classical mechanics to quantum mechanics.

In the dual thermostatics considered here, the counterpart of the mechanical action $I=\int L{\rm d}t$ is the entropy $S$. We will identify isoentropic submanifolds (of thermodynamical state space) with equipotential submanifolds (of mechanical state space). This is justified because, in the approach of emergence, {\it forces are (proportional to) entropy gradients}\/. In the particular case of the gravitational force, this identification has been put forward in ref. \cite{VERLINDE}; it coincides with the viewpoint applied in the theory of irreversibility \cite{PRIGOGINE} and, indeed, with the whole programme of the emergent physics paradigm. In this way the quantum mechanical exponential
\begin{equation}
\exp\left(-\frac{{\rm i}}{\hbar}I\right)
\label{droste}
\end{equation}
becomes, in the dual thermostatics,
\begin{equation}
\exp\left(\frac{S}{k_B}\right).
\label{quernel}
\end{equation}
The correspondence between expressions  (\ref{droste}) and (\ref{quernel}) has been known for long, having been discussed more recently in ref. \cite{BAEZ} from the point of view of statistical mechanics. However, we would like to stress that the theory being considered here as dual to quantum mechanics is {\it not}\/ statistical mechanics, but the thermostatics of equilibrium emerging from the latter.

{}Finally the connection between the mechanical time variable $t$ and the temperature $T$ is as follows:\footnote{This substitution is widely applied in thermal field theory, see {\it e.g.}\/ ref. \cite{TFT}.}
\begin{equation}
\frac{{\rm i}}{\hbar}t\longleftrightarrow-\frac{1}{k_BT},
\label{pelipell}
\end{equation}
where $\hbar$, $k_B$ are Planck's constant and Boltzmann's constant, respectively. The double arrow is to be understood as {\it replace every occurrence of ${\rm i}t/\hbar$ in the mechanical theory with $-1/k_BT$ in the thermostatical dual, and viceversa}\/. Quasistatic mechanics therefore corresponds to isothermal processes in the dual thermostatics.

\section{The quasistatic mechanics dual to thermostatics}

Given some specific thermostatical systems, below we illustrate how to define their corresponding (quasistatic) quantum mechanical duals.

\subsection{The ideal gas}

An expression for the entropy of a system in terms of its thermodynamical variables is called a {\it fundamental equation}\/ for the system \cite{CALLEN}. To be specific let us consider 1 mole of an ideal gas occupying a volume $V$ at a fixed temperature $T$. Its fundamental equation reads
\begin{equation}
S(V)=S_0+k_B\ln\left(\frac{V}{V_0}\right),
\label{kalen}
\end{equation}
where $S_0$ is the entropy in the fiducial state specified by $V_0$; we take $S_0$ to contain a constant contribution from the fixed temperature $T$. The entropy depends only on the volume $V$; the latter, running over $(0,\infty)$, can be regarded as the thermodynamical coordinate for the {\it isothermal}\/ processes of an ideal gas. 

In order to construct a kinetic energy operator $K$ for the quantum theory, the standard rule is
\begin{equation}
K:=-\frac{\hbar^2}{2M}\nabla^2,
\label{defkine}
\end{equation}
where $\nabla^2$ is the Laplacian operator on functions. By definition, the Laplacian requires a metric $g_{ij}$:
\begin{equation}
\nabla^2=\frac{1}{\sqrt{g}}\partial_i\left(\sqrt{g}g^{ik}\partial_k\right),\qquad g=\vert\det(g_{ij})\vert.
\label{niazzo}
\end{equation}
The fundamental equation (\ref{kalen}) provides us with a clue as to which metric can be meaningfully chosen. We first observe that Eq. (\ref{kalen}) is valid in 3--dimensional space, where the volume $V$ scales like $r^3$; here $r,\theta,\varphi$ are spherical coordinates. This suggests using the Euclidean metric in $\mathbb{R}^3$,
\begin{equation}
{\rm d}s^2={\rm d}r^2+r^2{\rm d}\theta^2+r^2\sin^2\theta{\rm d}\varphi^2,
\label{metrikk}
\end{equation}
and imposing the following two requirements. First, motion along the radial direction $r$ must cause an increase or decrease of the entropy, as per the fundamental equation (\ref{kalen}), with $V=4\pi r^3/3$; second, the sphere $r=r_0$ must define an isoentropic surface for each $r_0$. 

{}Further support for our argument follows from a classic result by H. Weyl:\footnote{We quote this result from ref. \cite{SINGER}: let $R\subset\mathbb{R}^3$ be a bounded region with piecewise smooth boundary, and let $V(R)=\int_R\sqrt{g}\,{\rm d}^3x$ denote its volume with respect to some Riemannian metric on $\mathbb{R}^3$. Then the eigenvalue equation for the Laplacian on $R$, $\nabla^2f=\lambda f$, supplemented with some mild boundary conditions, has a countable infinity of real eigenvalues $\lambda_n$ satisfying $0\geq\lambda_1\geq \lambda_2\geq \lambda_3\geq \ldots$. These eigenvalues can be arranged into a partition function $Z(t)$,
\begin{equation}
Z(t):={\rm Tr}\exp\left(t\nabla^2\right)=\sum_{n=1}^{\infty}\exp\left(t\lambda_n\right),
\label{partizione}
\end{equation}
and it turns out that the small $t$ asymptotics of $Z(t)$ is given by 
\begin{equation}
Z(t)\simeq \frac{V(R)}{(4\pi t)^{3/2}},\qquad t\to 0.
\label{totita}
\end{equation}
An analogous result holds within $\mathbb{R}^d$ (it is not necessary to assume that $d=3$; it is not necessary that the metric be the Euclidean one; it is also not necessary to assume that $R$ is a sphere). However, the Euclidean assumption is suggested by the fundamental equation (\ref{kalen}), while the assumption of spherical symmetry (in no way imposed by the ideal gas) provides a welcome simplification.} {\it the volume $V$ occupied by the ideal gas within Euclidean space is related, in a natural way, to the spectrum of the Laplacian operator within (and on the boundary surface of) $V$}\/.

We will initially define the Hilbert space ${\cal H}$ of {\it quasistatic quantum mechanics}\/ as the space of those states that minimise the expectation value of the kinetic energy, subject to the constraint that they be normalised (plus some boundary conditions to be specified below). Thus introducing a Lagrange multiplier $-\lambda\in\mathbb{R}$, we need to solve
\begin{equation}
\frac{\delta}{\delta\vert\psi\rangle}\left(\langle\psi\vert K\vert\psi\rangle-\lambda\langle\psi\vert\psi\rangle\right)=0, \qquad \langle\psi\vert\psi\rangle=1.
\label{lagranjj}
\end{equation}
Since $K$ is selfadjoint, Eq. (\ref{lagranjj}) leads to
\begin{equation}
K\vert\psi\rangle=\lambda\vert\psi\rangle,
\label{kpii}
\end{equation}
so the Hilbert space ${\cal H}$ is initially defined as
\begin{equation}
{\cal H}:={\rm Ker}\left(K-\lambda_{\rm min}\right),
\label{defdaf}
\end{equation}
where $\lambda_{\rm min}$ is the minimal kinetic energy; we have seen that $\lambda\geq 0$. We will presently see how the inclusion of a potential function $U$ affects the definition (\ref{defdaf}) of the Hilbert space.

\subsection{Motion along isoentropic surfaces}\label{zzatr}

We first analyse motion along a given isoentropic surface, which we take to be the unit sphere $S^2$. The angular part $\nabla^2_{S^2}$ of the Laplacian operator on $\mathbb{R}^3$ leads to the kinetic energy operator $K_{S^2}$:
\begin{equation}
K_{S^2}\psi:=-\frac{\hbar^2}{2M}\nabla^2_{S^2}\psi=-\frac{\hbar^2}{2M}\frac{1}{\sin\theta}\left[\frac{\partial}{\partial\theta}\left(\sin\theta\frac{\partial\psi}{\partial\theta}\right)
+\frac{1}{\sin\theta}\frac{\partial^2\psi}{\partial\varphi^2}\right].
\label{lanbaku}
\end{equation}
Within the space $L^2(S^2)$ the eigenvalues $\lambda$ of Eq. (\ref{kpii}) are $\hbar^2l(l+1)/(2M)$, with $l\in\mathbb{N}$ ; the least kinetic energy for motion on $S^2$ corresponds to the zeroth spherical harmonic $Y_{00}=(4\pi)^{-1/2}$:
\begin{equation}
K_{S^2}Y_{00}=0.
\label{banisin}
\end{equation}
The corresponding particle is completely delocalised on $S^2$, as befits the fact that its momentum vanishes exactly. The Hilbert space ${\cal H}_{S^2}$ is defined as the linear span of the spherical harmonic $Y_{00}$, {\it i.e.}\/,
\begin{equation}
{\cal H}_{S^2}={\rm Ker}\left(\nabla^2_{S^2}\right).
\label{dadade}
\end{equation}
On a compact, connected manifold, the only harmonic functions are the constants; the specific value $(4\pi)^{-1/2}$ is determined by normalisation. Although we have computed $\dim{\cal H}_{S^2}$ explicitly, the finite dimensionality of ${\rm Ker}\left(\nabla^2_{S^2}\right)\subset L^2(S^2)$ was already guaranteed on the basis of general results concerning the theory of elliptic operators on compact Riemannian manifolds \cite{SCHWARZ1}.\footnote{In this particular case, one can more simply apply the Hodge theorem \cite{SCHWARZ2}: since the 2--sphere $S^2$ is a compact, orientable Riemannian manifold, we have
$$
\dim{\rm Ker}\left(\nabla^2_{S^2}\right)=b^0(S^2)=1,
$$
where $b^0$ is the zeroth Betti number of the manifold in question.} A finite dimensional Hilbert space is a feature of many topological theories \cite{TOPO}: although a metric was initially required to define a Laplacian operator, the metric dependence is softened in the end, through the requirement of quasistatisticity (\ref{lagranjj}). 

{}Finally we can add a potential function $U=U(r)$ depending only on the radial variable $r$ and the previous arguments remain entirely valid. We then get back to the situation described in section \ref{greunda}: a particle moving quasistatically along the equipotential submanifolds of a certain potential.

\subsection{Motion across isoentropic surfaces}\label{izentro}

Next we analyse motion across isoentropic surfaces. The radial part $\nabla^2_r$ of the Laplacian operator on $\mathbb{R}^3$ gives rise to the kinetic energy operator $K_r$:
\begin{equation}
K_r\psi:=-\frac{\hbar^2}{2M}\nabla^2_{r}\psi=-\frac{\hbar^2}{2M}\left(\frac{{\rm d}^2\psi}{{\rm d}r^2}+\frac{2}{r}\frac{{\rm d}\psi}{{\rm d}r}\right).
\label{baku}
\end{equation}
By Eqs. (\ref{kpii}) and  (\ref{baku}) we need to solve
\begin{equation}
\frac{{\rm d}^2\psi}{{\rm d}r^2}+\frac{2}{r}\frac{{\rm d}\psi}{{\rm d}r}+c^2\psi=0,\qquad c^2:=\frac{2M\lambda}{\hbar^2}\geq 0;
\label{mifewe}
\end{equation}
a fundamental set of solutions is $\left\{\psi_{\pm}(r)=r^{-1}\exp(\pm {\rm i}cr)\right\}$. A vanishing kinetic energy is attained when $c=0$. However the corresponding wavefunction, $\psi(r)=1/r$, is neither regular at $r=0$, nor square integrable over the interval $(0,\infty)$. Imposing regularity of $\psi(r)$ at $r=0$ one is left with the wavefunctions
\begin{equation}
\psi(r)=\frac{1}{r}\sin\left(cr\right),
\label{eijjenbab}
\end{equation}
while the wavenumber $c\in\mathbb{R}$ remains undetermined. We can determine $c$ if we recall the relation between the squared wavefunction $\vert\psi\vert^2$ and the entropy \cite{VALENCIA}:
\begin{equation}
\vert\psi\vert^2=\exp\left(\frac{S}{k_B}\right).
\label{relationis}
\end{equation}
Collecting different microstates into a single pure quantum state is reminiscent of Von Neumann's density matrix formulation of the entropy of a mixed quantum state. However, even a pure state embodies a probability distribution; the latter has an associated Shannon entropy. The entropy of a pure state is not monotonic in time under Schroedinger evolution; this problem remains unsolved.

Let $r_0$ be the radius of the fiducial sphere in Eq. (\ref{kalen}). When evaluated at $r=r_0$, Eq. (\ref{relationis}) becomes, by Eq. (\ref{eijjenbab}),
\begin{equation}
\frac{1}{r_0}\sin(cr_0)=\exp\left(\frac{S_0}{2k_B}\right).
\label{sinsoluzion}
\end{equation}
Now the sine function is bounded between $-1$ and $+1$. This requires fine tuning the value of the fiducial entropy $S_0$ as a function of the fiducial radius $r_0$, or viceversa, if Eq. (\ref{sinsoluzion}) is to have a real solution for $c$. The simplest choice is to formally set $S_0=-\infty$. This choice has the added bonus that Eq. (\ref{sinsoluzion}) admits real solutions for $c$, without the need to fine tune $r_0$ as a function of $S_0$; it corresponds to imposing the additional boundary condition $\psi(r_0)=0$. Then the admissible eigenfunctions, with their corresponding wavenumbers $c_n\in\mathbb{R}$, are given by
\begin{equation}
\psi_n(r)=\sqrt{\frac{2}{r_0}}\,\frac{1}{r}\sin\left(c_nr\right),\qquad c_n=\frac{n\pi}{r_0} \qquad n=1,2,\ldots
\label{eijjen}
\end{equation}
We have normalised $\psi_n$ within $L^2\left([0,r_0]\right)$.

The least kinetic energy is attained when $n=1$. Therefore we define the Hilbert space ${\cal H}_r$ as the kernel
\begin{equation}
{\cal H}_r={\rm Ker}\left(\nabla^2_r+c_1^2\right).
\label{neell}
\end{equation}
This 1--dimensional space is generated by the wavefunction $\psi_1(r)$. More generally, the finite dimensionality of ${\rm Ker}\left(\nabla^2_r+c_n^2\right)\subset L^2([0,r_0])$ for all $n=1,2,\ldots$ is guaranteed by the theory of elliptic operators on compact Riemannian manifolds \cite{SCHWARZ1}. 

So far, the total Hilbert space ${\cal H}$ is the tensor product of the spaces (\ref{dadade}) and (\ref{neell}):
\begin{equation}
{\cal H}={\cal H}_{S^2}\otimes{\cal H}_r.
\label{lobin}
\end{equation}
We have up to now considered a free particle. If a potential function $U(r)$ is included, then the Hilbert space (\ref{neell}) must be redefined to be
\begin{equation}
{\cal H}_r={\rm Ker}\left(-\frac{\hbar^2}{2M}\nabla^2_r-\frac{\hbar^2}{2M}c_1^2+U(r)\right),
\label{neaputensia}
\end{equation}
and the latter substituted back into Eq. (\ref{lobin}). The above kernel remains finite dimensional. This is because the addition of $U(r)$ does not alter the ellipticity of the Hamiltonian, hence general theorems concerning the spectrum of elliptic operators on compact Riemannian manifolds continue to apply \cite{SCHWARZ1}. Of course, the presence of a potential on the quantum mechanical side modifies the fundamental equation (\ref{kalen}) of the corresponding thermostatics.

We close this section with some remarks.\\
{\it i)} The compact configuration space $[0,r_0]\times S^2$ has advantage that, due to energy quantisation, one can univocally identify a {\it nonvanishing}\/ state of least kinetic energy. On the noncompact configuration space $[0,\infty)\times S^2$, the allowed energy eigenvalues run over $[0,\infty)$, and no {\it nonvanishing}\/  state of least energy exists.\\
{\it ii)} Results analogous to those presented above would continue to hold if the free quantum particle were placed in a cubic  box of volume $L^3$, with vanishing boundary conditions for the wavefunction on the sides of the cube. The use of Cartesian coordinates renders isoentropic surfaces (now cubes) somewhat clumsier to work with than spheres, but the expectation value of the entropy (see Eq. (\ref{midfgh}) below) remains metric independent, and also the Hilbert space continues to be  1--dimensional.\\
{\it iii)}  Analogous results would hold as well if we worked in $d$--dimensional Euclidean space $\mathbb{R}^d$, {\it viz}\/: finite dimensionality of the Hilbert space, and metric independence of the expectation of the entropy.

\subsection{A metric free entropy}\label{miremire}

It is instructive to compute the expectation value of the entropy in the state (\ref{eijjen}). We set $V=4\pi r^3/3$, $V_0=4\pi r_0^3/3$, and write the quantum mechanical operator corresponding to the classical entropy of Eq. (\ref{kalen}) as
\begin{equation}
\hat S(r)=S_0+3k_B\ln\left(\frac{\hat r}{r_0}\right).
\label{ziele}
\end{equation}
The carets are meant to indicate quantum operators. Subtracting the infinite constant $S_0$ one finds an expectation value of the entropy
\begin{equation}
\langle\psi_n\vert \hat S\vert\psi_n\rangle=3k_B\int_0^{r_0}r^2\vert\psi_n(r)\vert^2\ln\left(\frac{r}{r_0}\right){\rm d}r
=3k_B\left(\frac{{\rm Si}(2\pi n)}{2\pi n}-1\right),
\label{midfgh}
\end{equation}
where ${\rm Si}(x):=\int_0^x t^{-1}\sin t\,{\rm d}t$ is the sine integral function. In particular, {\it all terms depending on $r_0$ drop out of Eq. (\ref{midfgh})}\/. This is in perfect agreement with the topological character \cite{TOPO} of our model: the entropy cannot depend on the radius $r_0$ of the fiducial sphere, because the latter requires a metric for its definition.

\subsection{The quantum mechanical partition function}

The quantum mechanical partition function $Z_{\rm qm}(t)$ is defined by
\begin{equation}
Z_{\rm qm}(t)=\sum_n{\rm dim}{\cal H}_n\,\exp\left(-\frac{{\rm i}}{\hbar}E_nt\right),
\label{hoff}
\end{equation}
where ${\cal H}_n$ is the Hilbert eigenspace corresponding to the energy eigenvalue $E_n$. The above sum is usually divergent, but it can be made to converge by Wick rotating the time variable as per
\begin{equation}
Z_{\rm qm}(\tau)=\sum_n{\rm dim}{\cal H}_n\,\exp\left(-\frac{1}{\hbar}E_n\tau\right).
\label{vicc}
\end{equation}
In the quasistatic limit, the above sum is dominated by the least energy eigenvalue, $E_{\rm min}$, and $Z_{\rm qm}(\tau)$ becomes $Z_{\rm qqm}(\tau)$, the subindex ``qqm" standing for {\it quasistatic quantum mechanics}\/: 
\begin{equation}
Z_{\rm qqm}(\tau)={\rm dim}{\cal H}_{\rm min}\,\exp\left(-\frac{1}{\hbar}E_{\rm min}\tau\right).
\label{domine}
\end{equation}
Therefore
\begin{equation}
Z_{\rm qqm}(0)={\rm dim}{\cal H}_{\rm min},
\label{loliloli}
\end{equation}
and {\it the partition function of quasistatic  quantum mechanics computes the dimension of the Hilbert space of quantum states}\/; also a conclusion that is  reminiscent of topological models \cite{TOPO}.

\section{Conclusions and outlook}\label{konkushon}

The application of differential and Riemannian geometry to the theory of thermodynamical fluctuations has turned out to be extremely useful \cite{RUPPEINER, BRAVETTI, NOIDUE}. Thus, {\it e.g.}\/, the classical thermodynamics of irreversible processes \cite{ONSAGER, PRIGOGINE} requires for its formulation a metric on phase space. This metric is provided by Onsager's matrix of kinetic coefficients $L_{ij}$. The metric enters the quantum mechanical dual theory \cite{VALENCIA} through the kinetic term in the mechanical Lagrangian.

On the contrary, the thermostatics of equilibrium processes \cite{CALLEN} is genuinely metric free. Therefore, if thermostatics is to possess any quantum mechanical dual at all, this dual theory should be a topological theory \cite{TOPO}, in the sense that it should be metric independent. 

That the classical thermostatics of equilibrium processes should possess a quantum mechanical dual is suggested by two observations. First, by the claim that quantum mechanics is an emergent phenomenon \cite{ELZE1, ELZE2, THOOFT, VALENCIA, SMOLIN, GALLEGO}. Second, by the widespread opinion that thermodynamics (be it of equilibrium \cite{CALLEN} or nonequilibrium \cite{ONSAGER, PRIGOGINE}) is the paradigm of all emergent sciences. These conclusions remain unaltered even if, as argued in ref. \cite{CALMET}, the emergent aspects of quantum mechanics can only become visible at very high energies.

Two guiding principles are at work here: the notion that forces are entropy gradients, and the requirement that all processes be quasistatic. Entropy gradients, while defining a direction for evolution, ignore microscopic structures, retaining only coarse grained averages: this is a feature of emergent phenomena. Ignoring the metric structure of the underlying manifold amounts to ignoring the kinetic term in the Lagrangian. Quantum mechanically, due to the uncertainty principle, the effects of the kinetic term cannot be cancelled completely, unless one accepts a complete delocalisation of the particle in space. The result of following these two guiding principles is a quasistatic quantum mechanics, which is dual to the classical thermostatics of equilibrium processes, and shares a number of key properties in common with topological, {\it i.e.}\/, metric free models.

After completion of this work there appeared ref. \cite{CODESIDO}, where the WKB expansion of quantum mechanics is developed from the point of view of topological string theory \cite{MARINO}. Ref. \cite{CODESIDO} provides further evidence of the existing links between topological theories and quantum mechanics. Some of these links have been analysed in the present paper, from the alternative standpoint of the approach of emergence of quantum theory; further connections are being studied in an upcoming publication \cite{UPCOMING}.

\vskip.5cm
\noindent
{\bf Acknowledgements} Research supported by grant no. ENE2015-71333-R (Spain).

\end{document}